# Manipulating Hubbard-type Coulomb blockade effect of metallic wires embedded in an insulator

Xing Yang[1], Huimin Wang[2], Jing-Jing Xian[1], Sheng Meng[2], Naoto Nagaosa[3,4], Wen-Hao Zhang[1], Hai-Wen Liu[5], Zi-Heng Ling[1], Kai Fan[1], Zhi-Mo Zhang[1], Le Qin[1], Zhi-Hao Zhang[1], Yan Liang[1], Ying-Shuang Fu[1*]

[1]School of Physics and Wuhan National High Magnetic Field Center, Huazhong University of Science and Technology, Wuhan 430074, China

[2]Beijing National Laboratory for Condensed Matter Physics and Institute of Physics, Chinese Academy of Sciences, Beijing 100190, China

[3]RIKEN Center for Emergent Matter Science (CEMS), Wako, Saitama 351-0198, Japan

[4]Department of Applied Physics, University of Tokyo, Tokyo 113-8656, Japan

[5]Center for Advanced Quantum Studies, Department of Physics, Beijing Normal University, Beijing 100875, China

*yfu@hust.edu.cn

**Correlated states emerge in low-dimensional systems owing to enhanced Coulomb interactions. Elucidating these states requires atomic scale characterization and delicate control capabilities. In this study, spectroscopic imaging-scanning tunneling microscopy was employed to investigate the correlated states residing in the one-dimensional electrons of the monolayer and bilayer $MoSe_2$ mirror twin boundary (MTB). The Coulomb energies, determined by the wire length, drive the MTB into two types of ground states with distinct respective out-of-phase and in-phase charge orders. The two ground states can be reversibly converted through a metastable zero-energy state with *in situ* voltage pulses, which tunes the electron filling of the MTB via a polaronic process, as substantiated by first-principles calculations. Our modified Hubbard model reveals the ground states as correlated insulators from an on-site *U*-originated Coulomb interaction, dubbed Hubbard-type Coulomb blockade effect. Our**

**work sets a foundation for understanding correlated physics in complex systems and for tailoring quantum states for nano-electronics applications.**

In reduced dimensions, the screening effect is suppressed, resulting in enhanced Coulomb interactions. Owing to their structural simplicity, low-dimensional systems provide model platforms for studying correlated physics [1-5]. In particular, electrons in one dimension (1D) are endowed with divergent electron susceptibility on account of their perfect Fermi surface nesting, which makes them susceptible to interactions [6,7]. They are exemplified by the Peierls-type charge density wave (CDW) originating from the periodic lattice distortion induced by the electron–phonon interaction [8-10]. They are also exemplified by the fractionalized spin-charge separation in the Tomonaga–Luttinger liquid (TLL) as a result of electron–electron interactions [11-12]. Thus, 1D electrons serve as a paradigmatic system for investigations of emergent correlated states.

Actual 1D systems are subject to quantum fluctuations, which profoundly influence the correlated states [13-15]. Inter-chain coupling can suppress such influence; however, it also introduces complications to the system. Recent studies have identified a new 1D metallic system residing in the mirror twin boundary (MTB) of transition metal dichacogenides [16]. This system is devoid of inter-chain coupling and has negligible quantum fluctuations because of its embedment inside the bulk insulating matrix [15,17]. These properties render it an ideal candidate for studying the correlated physics in 1D.

In correlated phases, the electron filling adjustment is not only decisive for pinning down the nature of the ground states, but also engages with Coulomb interactions for tuning phase transitions. It is desirable to investigate the evolution of the correlated phases with the *in-situ* response of electron filling. Conventional methods of carrier doping inevitably introduce disorder to the systems. In low-dimensional systems, electric gating acts as another means of tuning the electron filling. Recent achievements in such phase transitions among Mott-like insulators, superconductors, and topological orders have been realized in twisted magic-angle graphene superlattices [18-20]. The electric gating method, however, requires sophisticated procedures for device fabrication.

We herein characterize the correlated states in 1D MTB wires and their *in-situ* manipulation with spectroscopic imaging scanning tunneling microscopy (SI-STM) at 4 K. One-dimensional wires of a finite length are found to be driven into two types of correlated insulating ground states that arise from a dubbed Hubbard-type Coulomb blockade (CB) effect. With the application of STM voltage pulses, the electron filling of the MTB is tuned, switching the two ground states reversibly in a controlled manner. Our study clarifies the nature of the correlated insulating ground states and demonstrates the feasibility of locally controlling the correlated states of individual wires, which establishes a model platform for emulating the Hubbard model and tailoring quantum states.

Figure 1(a) shows the topography of the monolayer and bilayer $MoSe_2$ films grown on graphene-covered SiC substrate (Supplementary Methods). Bright straight-line defects are MTBs embedded in the films. Their crystal structure [Fig. 1(b), inset] conforms to our atomic-resolution imaging (Supplementary Note 1). Line spectra along the wire show quantized discrete levels [Figs. 1(c) and (d)], which are quantum well states (QWSs) due to confinement of electrons propagating along the finite MTB wire (Supplementary Note 2). The tunneling spectrum of the MTB [Fig. 1(b)] features an abrupt conductance onset peak at 0.27 eV, a series of peaks related to the QWSs, and a spectral dip at −0.65 eV. These spectroscopic features are effectively captured by our density functional theory (DFT) calculations, with a Mo-Mo bond length of 2.66 Å at the MTB. The measured lattice constant along the MTB is 0.34 nm, agreeing with the DFT calculation (0.33 nm).

Low-energy states around the Fermi level $E_F$ [Figs. 2(a,b)] exhibit fully gapped states and strong peaks at the gap edges [Figs. 2(a,b), red arrows] with evidently higher intensities than the QWSs. Satellite phonon peaks coexist for both the QWSs and the low-energy peaks with an energy spacing of 14±1 meV [Figs. 2(a,b), blue arrows] [17]. Interestingly, two types of low-energy states are observed at different MTBs, featuring different categories of phase relations. Specifically, for the type 1 MTB [Fig. 2(a)], its conductance modulations of the two gap edges are approximately anti-phase. Its spectral gap size, $E_{g0}$, is uniform throughout the wire, whose magnitude is much larger than the energy spacing between QWSs, $\Delta E_{QWS}$. The type 2 MTB [Fig. 2(b)] has a similar spectral gap as type 1, but with substantial differences. First, its gap size, $E_{g1}$, is close to $\Delta E_{QWS}$. Second, the spatial modulation of the gap edges is maintained strictly in-phase. Third, the two second-nearest peaks to $E_F$ [Fig. 2(b), purple arrows] are even more enhanced. We define the energy separation between the first occupied peak and the second unoccupied peak near $E_F$ as $E_{g2}$. Hereafter, the low-energy states of the type 1 and type 2 MTBs are equivalently designated with their phase relations as the out-of-phase state and in-phase state, respectively.

Statistics over 80 MTBs indicate that the ratio between the type 1 and type 2 wires is approximately 4:1. Moreover, the gap sizes of both low-energy states show a pertinent inversely proportional relation with the wire length, *L* [Figs. 2(c,d)] [21,22]. A similar relation exists between $\Delta E_{QWS}$ and *L*, which prompts us to evaluate their relations. For the type 1 MTBs with various lengths, their ratios of $E_{g0}/\Delta E_{QWS}$ are similar, with statistics of 2.25±0.14 (standard deviation) [Fig. 2(e)]. Similar ratios of $E_{g2}/E_{g1}$ apply to the type 2 MTBs with statistics of $1.68 \pm 0.07$ [Fig. 2(f)]. Both types of MTBs in the monolayer and double-layer $MoSe_2$ exhibit an identical dependence of the spectral gap size with the wire length. Moreover, no difference was found between the graphene substrate and highly oriented pyrolytic graphite (HOPG) [Figs. 2(c–f)].

These observations suggest that the low-energy states and the QWSs are implicitly related, which are thereafter all equally treated as discrete levels by scrutinizing their numbers of nodes. In the type 1 MTBs, the node number differs by 1 for adjacent levels [Fig. 3(a)]. In the type 2 MTBs, similar node-increment rules apply, except for the low-energy levels around $E_F$, which have the same node number [Fig. 3(c)]. This notion has several implications. First, the low energy levels may have the same origin as the QWSs. Second, the two gap-edge levels of the in-phase state could be essentially from the same QWS, which is gapped by spin degeneracy lifting via a Mott-like insulating mechanism, like Ref. [18]. Third, the Coulomb interaction should exist in the type 1 MTBs as well, which evokes the possible gap opening of the out-of-phase state.

The above implications can be examined with an electron filling adjustment, for which we expect an effect of the Coulomb interaction. Fig. 3(a) presents a conductance plot along an MTB in out-of-phase state. The out-of-phase relation of the two gap-edge levels gradually diminishes upon approaching the wire ends. By applying tip pulses of 1.9 V, all the discrete levels rigidly shift toward higher energy. The spectral gap of the out-of-phase state disappears immediately upon the upper gap-edge level partially overlapping with $E_F$ [Fig. 3(b)]. In this new state, the dubbed zero-energy state, all the discrete levels become equidistant in energy with the node numbers differing by 1 for the adjacent levels. This zero-energy state can be transformed into the in-phase state with further pulsing [Fig. 3(c)]. For the in-phase state, the modulation of the two gap-edge levels is maintained strictly in-phase throughout the entire MTB. In many cases, the zero-energy state can spontaneously transform into the in-phase state or the out-of-phase state during spectroscopic measurements, demonstrating it is unstable. More interestingly, such phase transitions are fully reversible with a series of opposite polarity pulses of −1.9 V (Supplementary Note 3).

For the MTB in different ground states, its overall electronic structure is preserved [Fig. 3(d), Fig. S4]. Merely $E_F$ is tuned by the voltage pulses, thereby demonstrating that the MTB accommodates additional charges. The additional charge is not trapped by localized defects below the graphene [23], which would otherwise cause electronic inhomogeneity along the MTB. To understand the charging mechanism, DFT calculations were carried out, with one additional electron added per six MTB units. As shown in Fig. 3(e), the band structure of the MTB is reserved with an $E_F$ increase of 81 meV. Concomitantly, the MTB Se-Se (Mo-Mo) bond length slightly increases by 0.024 pm (0.0073 pm) [Fig. 3(f)], and the added charges are distributed mainly on the Mo atoms of the MTB [Fig. 3(g)] as well as on the edges of the constructed simulation structure [Fig. S5]. This results in a uniform $E_F$ adjustment across the wire. Hence, the collective structural relaxation of the MTB bonds stabilizes the injected charge through a polaronic process [24].

Hereafter, we elucidate the ground states of the MTBs. The Peierls-type CDW state, as reported in Ref. 17, could be excluded, because CDW gap should be robust against electron filling tuning, contrasting with our observation. The classical CB effect could display a Coulomb gap and is sensitive to electron filling [25]. The MTB can be considered an elongated quantum dot, whose capacitance is proportional to the wire length. However, this is unlikely either. First, for the monolayer MTB of a given length, its capacitance relative to the graphene substrate should be approximately twice of that in the bilayer films, which contradicts the identical spectral gap sizes for both state types. Second, the MTBs grown on a HOPG substrate, which should have a different capacitance, quantitatively show the same length dependence of spectral gap as those on graphene [Figs. 2(c–f)]. Furthermore, the spectral gap sizes are invariant against changing tip-sample separations as well, where the capacitance between the tip and MTB varies [Fig. S6]. Third, the pulsing application on a half-covered MTB (Fig. S7) and the two contacting MTBs (Fig. S8) can only switch the ground state of the partial MTB while leaving the other half intact. These results suggest that the Coulomb interaction is local in nature. Otherwise, long-range Coulomb interactions would mutually affect the ground states of the two contacting quantum dots. In this regard, the MTB ground states are ascribed as correlated insulators.

The three states outlined in Figs. 3(a–c) all have the same origin as that induced by QWSs mediated by Coulomb interactions.

To quantitatively depict such ascription, we consider the effect of Hubbard $U$ on the ground state [26] and model the MTB with a finite chain with $N+1$ atoms and a unity lattice constant. Confinement from the finite chain results in discrete levels, whose wave functions are labeled by momentum $k$ with the fixed boundary condition. We start with the Hubbard Hamiltonian given by

$$H = -\sum_{i,\sigma}[t(c_{i,\sigma}^\dagger c_{i+1,\sigma} + c_{i+1,\sigma}^\dagger c_{i,\sigma}) - \mu n_{i,\sigma}] + \frac{U}{2}\sum_i (n_i)^2 \qquad (1)$$

where $\mu$ is the chemical potential, $n_{i,\sigma} = c_{i,\sigma}^\dagger c_{i,\sigma}$ is electron number of the spin $\sigma =\uparrow,\downarrow$ at site $i$, and $n_i = \sum_\sigma n_{i,\sigma}$. The eigenstate of the first term is $\Psi_k(i) = \sqrt{\frac{2}{N+1}}\sin\left(\frac{\pi k}{N+1}i\right)$. Then putting $n_{i,\sigma} = \sum_{k,k'} \Psi_k(i)\Psi_{k'}(i)\, c_{k,\sigma}^\dagger c_{k',\sigma}$ into the second term in Eq. (1) and neglecting the Fock term, one obtains the effective Coulomb interaction

$$H_U = \sum_k u_1 n_{k,\uparrow} n_{k,\downarrow} + \frac{1}{2}\sum_{k\neq k'} u_2\, n_k n_{k'} \qquad (2)$$

with $u_1 = \frac{3U}{2(N+1)}$, and $u_2 = \frac{U}{N+1}$.

Here, only the Coulomb interaction between two one-electron states, namely the highest occupied level $k_0$ and an unoccupied level $k_t$ hosting the tunneling electron, is considered, because merely those two states are relevant at low energies. Thus, Eq. (2) is written as

$$H_U = \begin{cases} u_1 n_{k_0,\uparrow} n_{k_0,\downarrow} & k_t = k_0 \\ u_2 n_{k_0} n_{k_t} & k_t \neq k_0 \end{cases} \quad (3)$$

Therefore, when the $k_0$ state is doubly occupied, corresponding to the type 1 MTB, an electron tunneling into any $k_t$ state will feel the Coulomb interaction of two electrons. The spectral gap size should be the sum of the quantization energy and the Coulomb energy, i.e. $E_{g0} = \Delta E_{QWS} + \frac{2U}{N+1}$ [Fig. 4(a)]. Similarly, for the case of singly occupied $k_0$ state, corresponding to the type 2 MTB, an electron tunneling into any $k_t$ state just feels the Coulomb interaction of one electron. But, its strength relies on the specific $k_t$ state. In the case of tunneling into the lowest unoccupied level, namely $k_t = k_0$, the spectral gap size is $E_{g1} = \frac{3U}{2(N+1)}$. And, when tunneling into the second lowest unoccupied level, namely $k_t \neq k_0$, the spectral gap is $E_{g2} = \Delta E_{QWS} + \frac{U}{N+1}$ [Fig. 4(c)]. On the other hand, if the $k_0$ state is right at $E_F$, charge fluctuations effectively screen the Coulomb interaction, removing the correlated gap [Fig. 4(b), Supplementary Note 6].

With the above modeling analysis, we determine $U$ with experimentally extracted parameters. Under a linear energy dispersion, we have $\Delta E_{QWS} = \frac{E_Q}{L}$, where $L = (N+1)a_0$ with the MTB unit $a_0$, and $E_Q$ is experimentally determined as 0.74±0.02 eV·nm (Supplementary Note 7). For the in-phase state, comparison of the experimental [shown in Fig. 2(d)] and theoretical $E_{g1}$ yields $U = 1.9\pm0.1$ eV. Similar comparison for the out-of-phase state determines a similar $U = 1.4 \pm 0.2$ eV, considering the simplicity of our model. The theory also gives $\frac{E_{g2}}{E_{g1}} = 1.43$ and $\frac{E_{g0}}{\Delta E_{QWS}} = 2.22$, conforming to the measured respective values of 1.68±0.07 and 2.25±0.14 [Figs. 2(e and f)].

As shown in Eq. (2), the electron confinement of wave functions renders the effective Coulomb interaction normalized by the chain site $N$, which becomes the on-site $U$ at the single-site limit. The effective Coulomb energy is remarkably analogous to the charging energy $e^2/(2C)$ in the classical CB effect, where the capacitance $C$ is proportional to the wire length. Thus, its dictation on the correlated insulating state can be deemed a Hubbard-type CB effect. The on-site $U$ suggests TLL behavior in the MTB [21,22], as indeed observed from Fig. S11.

In summary, we have discovered an unprecedented Hubbard-type CB effect, which could establish an alternative platform for emulating the Hubbard model in artificial quantum systems, as recently demonstrated with bosons [27] or fermions [28]. The Hubbard-type CB effect does not explicitly require inputs from the specific system of MoSe$_2$ MTB, which is beyond the scope of clarifying the controversies over its ground state. Such effect should be generally applicable to low dimensional metallic systems embedded in insulating environments, where the dielectric insulating matrix effectively reduces the Coulomb strength. Our strategy of *in-situ* electron-filling adjustment allows the addressing and manipulating of individual nano-objects with atomic-scale precision. This may foster the understanding of correlation physics in complex compounds, and the tailoring of quantum states for nano-electronics applications.

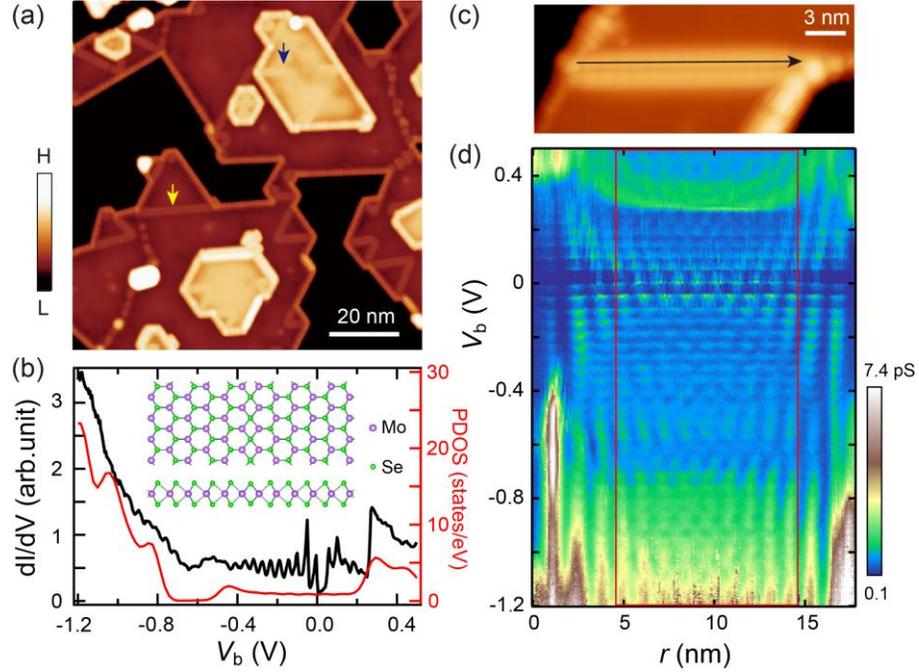

Figure 1. Topography and electronic structure of MTB on MoSe$_2$. (a) STM image ($V_t$ = 500 mV, $I_t$ = 20 pA) of MTB on MoSe$_2$. Exemplified monolayer and bilayer MTBs are marked with yellow and black arrows, respectively. (b) DFT-calculated DOS (red curve) and experimental spectrum (black curve) of an MTB. Inset: Top view and side view of the MTB crystal structure. (c) Magnified STM image ($V_t$ = 500 mV, $I_t$ = 10 pA) of a monolayer MTB. (d) Conductance plot ($V_t$ = 500 mV, $I_t$ = 100 pA, $V_{mod}$ = 3.54mV) obtained along the black line in (c). The spectrum in (b) is averaged from the spectra of (d) in rectangle.

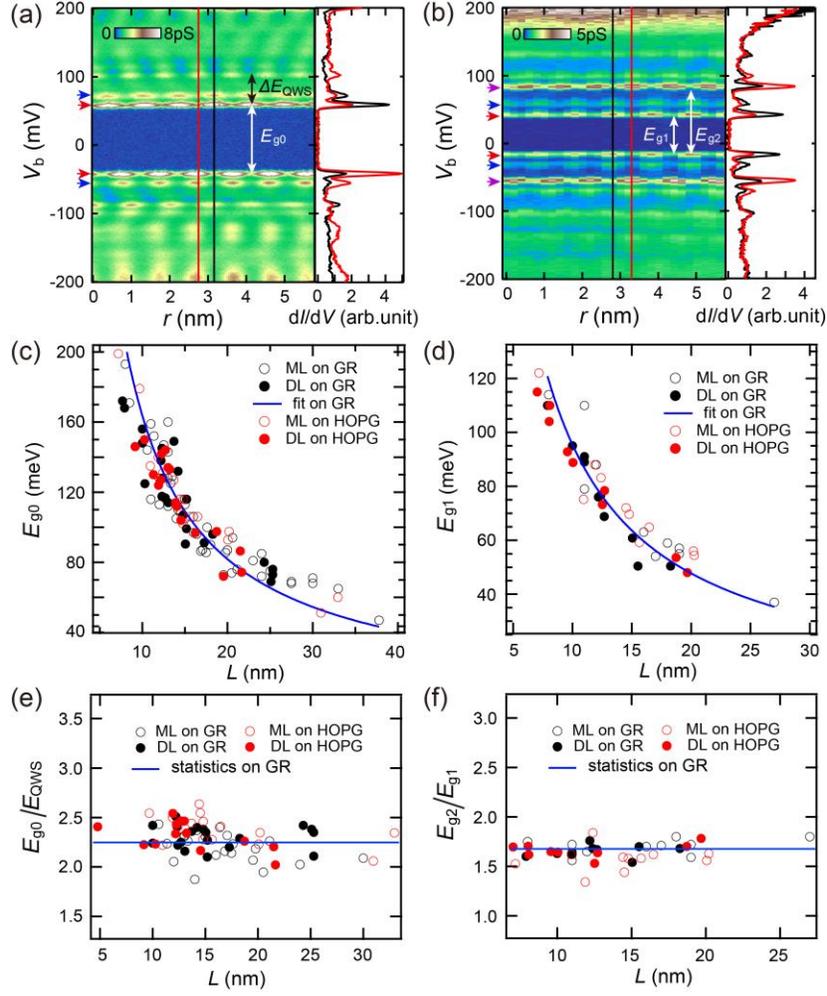

Figure 2. Two types of low-energy ground states of MTB. (a,b) Left: 2D conductance plot [$V_t = 200$ mV, $I_t = 100$ pA, $V_{mod}=1.414$ mV (rms)] obtained along an MTB showing QWSs and out-of-phase (a) [in-phase (b)] ground state. Right: The point spectra are extracted along two conductance modulation extrema from the red and black lines. (c,d) Statistics showing out-of-phase gap ($E_{g0}$) [in-phase gap ($E_{g1}$)] as a function of wire length ($L$). The blue curves are fittings to the data on graphene (GR) with an inverse proportional relation, yielding $E_{g0} = \dfrac{1.64 \pm 0.04 (\text{eV} \cdot \text{nm})}{L}$ and $E_{g1} = \dfrac{0.95 \pm 0.03 (\text{eV} \cdot \text{nm})}{L}$. (e,f) Statistics showing

the ratios of $E_{g0}/\Delta E_{QWS}$ and $E_{g2}/E_{g1}$, respectively, whose statistical values on the graphene substrate are marked with blue lines. In (c–f), the open (solid) dots mark the corresponding values measured on monolayer (ML) [double layer (DL)] MTBs, with the black (red) color representing the data on the graphene (HOPG) substrate.

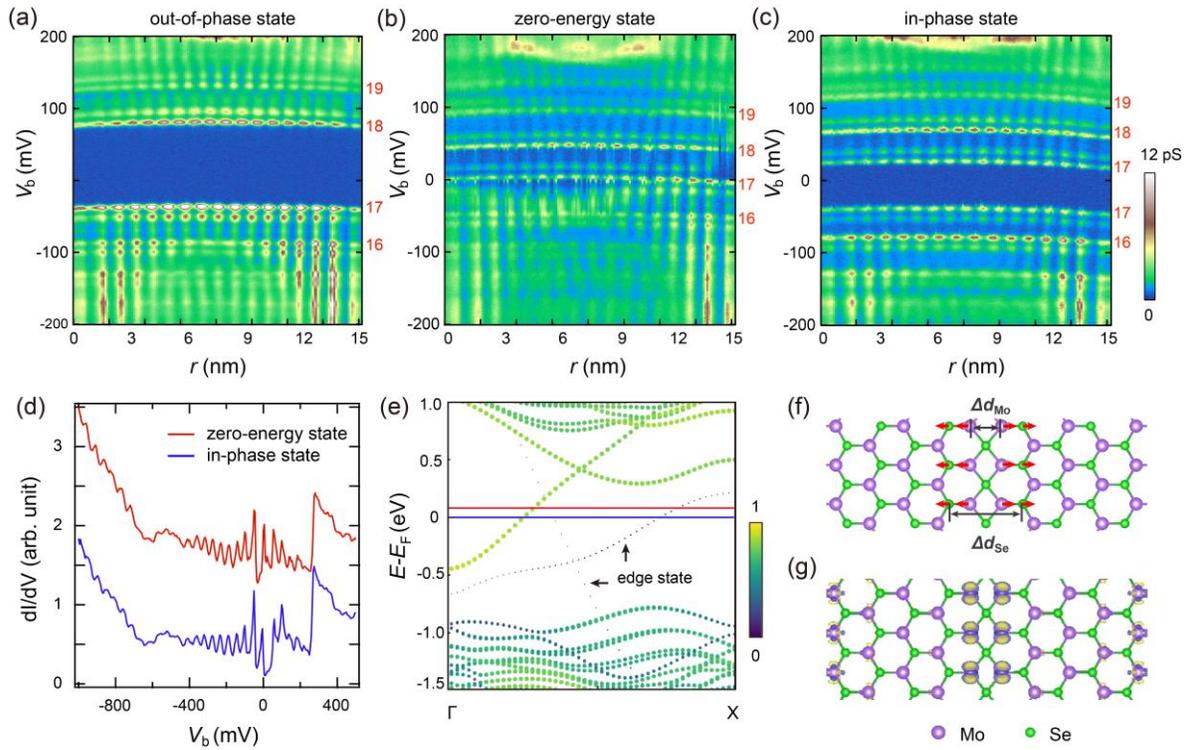

Figure 3. Ground-state transitions of MTB manipulated with tip pulses. (a–c) 2D conductance plot [$V_t$ = 200 mV, $I_t$ = 100 pA, $V_{mod}$ = 1.414 mV (rms)] of the same MTB shown in Fig. 1(c), displaying different ground states. (d) Large energy scale spectra [$V_t$ = 500 mV, $I_t$ = 100 pA, $V_{mod}$ = 3.54 mV (rms)] of the MTB in different states. The spectra have been offset vertically for clarity. (e) MTB calculated bands. The red (blue) horizontal

line depicts $E_F$ with (without) charged electrons. The color bar depicts the spectral weight of the state density on the MTB. Black arrows mark the edge states of the constructed simulation structure. The $X$ point corresponds to 0.95 Å$^{-1}$. (f) Optimized atomic structure of MTB with charged electrons. The lattice relaxation directions caused by charging are indicated with red arrows. (g) Calculated differential charge distribution of the charged relative to the uncharged MTB. The positive (negative) charge-density difference is represented by yellow (blue) areas. The isosurface value is $1\times10^{-3}$ electrons/Å$^3$.

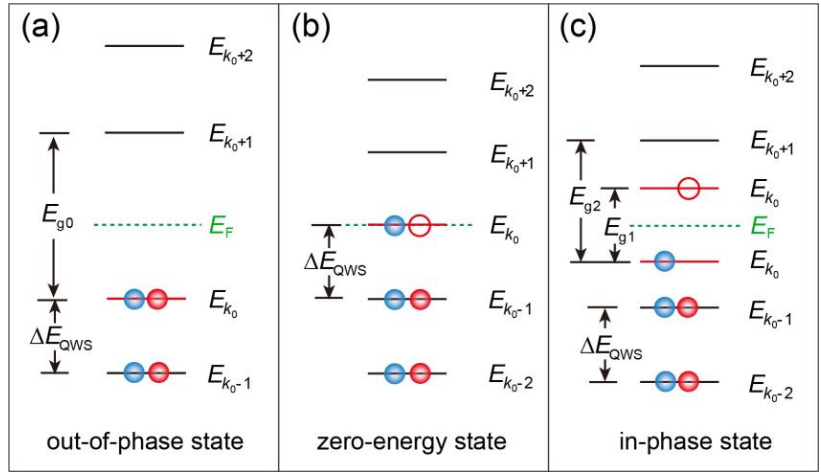

Figure 4. Hubbard model of the ground state of MTB. Schematics showing an energy level diagram of the correlated states of the MTB, namely, out-of-phase state (a), zero-energy state (b), and in-phase state (c), respectively. Each discrete level is marked with its wave vector, and $E_{k0}$ level is highlighted with a red line. The spin-up (spin-down) electrons are depicted with red (blue) balls. The solid (hollow) balls represent electrons residing occupied (unoccupied) levels.

**Acknowledgement**: We thank Gang Li, Tetsuo Hanaguri, Jing-Tao Lü, and Ying-Hai Wu for discussions. This work was funded by the National Key Research and Development Program of China (Grant Nos. 2017YFA0403501 and 2016YFA0401003), the National Science Foundation of China (Grant Nos. 11874161, 11522431, 11474112, 11774105, and 21873033), JST CREST Grant Nos. JPMJCR1874 and JPMJCR16F1, and JSPS KAKENHI Grant Nos. 18H03676 and 26103006.